\documentclass[showpacs,aps,pre,twocolumn]{revtex4-1}
\pdfoutput=1

\usepackage{cmap}
\usepackage{graphicx}
\usepackage{fncylab,environ}
\usepackage{url}

\usepackage{amssymb}
\usepackage{amsmath}

\usepackage{tikz}
\usepackage{pgfplots}

\usepackage{hyperref}

\usetikzlibrary{external}

\newcommand{\trc}{\operatorname{tr}}

\labelformat{figure}{Figure #1}
\labelformat{equation}{(#1)}
\labelformat{section}{Section #1}
\labelformat{table}{Table #1}

\newcommand\nll{
      \mathrel{\mkern2.1mu\not\mkern-2.1mu\ll}}

\makeatletter

\begin{document}

\title{Marginal stability in jammed packings: quasicontacts and weak contacts}

\author{Yoav Kallus}
\affiliation{Princeton Center for Theoretical Science, Princeton University, Princeton, New Jersey 08544}

\author{Salvatore Torquato}
\affiliation{Department of Physics, Princeton University, Princeton, New Jersey 08544}
\affiliation{Department of Chemistry, Princeton University, Princeton, New Jersey 08544}
\affiliation{Program in Applied and Computational Mathematics, Princeton University, Princeton, New Jersey 08544}
\affiliation{Princeton Institute of the Science and Technology of Materials, Princeton University, Princeton, New Jersey 08544}

\date{\today}

\begin{abstract}
    Maximally random jammed (MRJ) sphere packing is a prototypical example
    of a system naturally poised at the margin between underconstraint and
    overconstraint. This marginal stability has traditionally been understood
    in terms of isostaticity, the equality of the number of mechanical
    contacts and the number of degrees of freedom. Quasicontacts, pairs of spheres on the verge of
    coming in contact, are irrelevant for static stability, but they come into play when
    considering dynamic stability, as does the distribution of contact forces.
    We show that the effects of marginal dynamic stability, as manifested in
    the distributions of quasicontacts and weak contacts, are consequential
    and nontrivial. We study these ideas first in the context of MRJ packing of $d$-dimensional
    spheres, where we show that the abundance of quasicontacts grows at a faster
    rate than that of contacts. We reexamine a calculation of Jin et al.\ 
    (Phys. Rev. E \textbf{82}, 051126, 2010), where quasicontacts were originally
    neglected, and we explore the effect of their inclusion in the calculation. This
    analysis yields an estimate of the asymptotic behavior of the packing density
    in high dimensions. We argue that this estimate should be reinterpreted as a
    lower bound. The latter part of the paper is devoted to Bravais lattice
    packings that possess the minimum number of contacts to maintain mechanical stability.
    We show that quasicontacts play an even more important role in these packings.
    We also show that jammed lattices are a useful setting for studying the
    Edwards ensemble, which weights each mechanically stable configuration equally
    and does not account for dynamics. This \textit{ansatz} fails to predict the power-law
    distribution of near-zero contact forces, $P(f)\sim f^\theta$.
\end{abstract}
\pacs{61.20.-p, 05.65.+b, 61.50.Ah}

\maketitle

\section{Introduction}\label{sec:intro}

The margin between underconstraint and overconstraint in the stability of disordered systems
is associated with many unusual and not completely understood phenomena, such as glass
formation in covalently bonded materials \cite{glassconstraint}, peaks in computational complexity
in NP-hard problems \cite{satunsat}, and critical
heterogeneity in mechanical response of elastic networks \cite{fibre}.
Randomly packed frictionless hard spheres at the jamming transition are a prototypical example of
a system that is driven from the region of underconstraint just to the edge
of overconstraint. Maximally random jammed (MRJ) packings are those that exhibit the
least order, as measured by a variety of different order metrics, from the set
of all strictly jammed (mechanically stable) packings \cite{torqmrj}. Such packings
are naturally poised at the critical margin without the need to tune parameters.
The idea of marginal constraint in MRJ configurations
has long been synonymous with \textit{isostaticity}, the equality between the number
of mechanical contacts between spheres and the number of translational degrees
of freedom \cite{torqiso}. Recently, however, it has been argued that in addition to marginal
constraint from the point of view of mechanical stability, MRJ configurations
might also exhibit marginal constraint from the point of view of dynamic
stability \cite{wyart,lerner}. Mechanical stability reflects the impossibility of reducing
the volume occupied by the system through a continuous motion of its
particles such that the volume does not increase at any time during the motion \cite{donevlp}.
Dynamic stability reflects the impossibility of reducing
the volume even through motions that cause small rearrangements in the contact network,
and which might at first increase the volume. Wyart showed that
marginal dynamic stability is associated with power-law distributions of
the gaps between pairs of spheres that are \textit{nearly} in contact -- called quasicontacts --
and of the forces between pairs of spheres that \textit{are} in contact, but exert a nearly zero
contact force on each other \cite{wyart}.

In Sections II--III, we will consider MRJ configurations of frictionless hard spheres in
a Euclidean space of $d$ dimensions and study the abundance of quasicontacts.
Recently, many works have been devoted to identifying the asymptotic scaling
behavior of various densities associated with MRJ packings and the prejamming
dynamics of hard-sphere fluids in the limit $d\to\infty$, and this question
has lead to many advances in the theory of jamming and the glass transition
\cite{charbcorwin,charbikeda,parisi10,jin,ikeda,charbzamp}.
We will show that the abundance of quasicontact,
required for reasons of dynamic stability, imposes strong structural constraints
that are as important as those imposed by the network of mechanical contacts.
The power-law singularities of active constraints on the verge of becoming inactive
and \textit{vice versa} are not unique to MRJ packings, and they arise in other
disordered systems \cite{wyart-2014}. In Sections IV--V, we analyze jammed
Bravais lattice sphere packings, which are closely related to MRJ packings.

The dimensional dependence of the number of mechanical contacts around an
average sphere can be easily derived from Maxwellian constraint
counting: since each contact is incident on two spheres and imposes one constraint, and each sphere has
$d$ translational degrees of freedom, each sphere should have, on average, $2d$
contacts. Quasicontacts, on the other hand, 
do not contribute to mechanical stability, but
clearly participate in the trapping behavior as a configuration dynamically approaches
the jamming point. These contacts show up as a power-law divergence of the pair-correlation
function near the contact distance (historically, this was known as the ``square-root divergence,''
but the power-law exponent is no longer agreed to be 1/2) \cite{donev05,silbert06,charbcorwin}.
Quasicontacts are challenging to study, because unlike mechanical
contacts, identifying specific pairs as quasicontacts requires the definition of
an arbitrary cutoff distance \cite{jordey} or another criterion, such as 
adjacency of Voronoi regions \cite{KumarVor}. 
Since this divergent part of the pair correlation overlaps with the well-behaved part,
we do not attempt to identify specific pairs as quasicontacts and only study 
the influence of the abundance of pairs at small separations.

\section{Quasicontacts in disordered packings}\label{sec:mrj}

We would like to understand the characteristics of the quasicontact and weak contact
distributions as a function of dimension.
The calculation of Ref.\ \cite{wyart} is set in a fixed number of dimensions but
applies equally well in any dimension. We must
only reintroduce the dimensional dependence that was originally left out in order to determine
the expected number of quasicontacts around the average sphere. 
Consider an MRJ packing of $N$ unit-diameter spheres with centers $\mathbf{r}_k$, $k=1,\ldots, N$.
Isostaticity guarantees that for any contact $(i,j)$, 
there exists a continuous motion $\delta\mathbf{r}_k(s)$
of the sphere centers such that the distance between the centers of the spheres $i$ and $j$
is $1+s$, and the distance between the centers of any other pair originally in contact remains $1$.
In the process of this motion, the work performed by the packing is given by
\begin{equation}
    \begin{aligned}
	p\delta V &= f_{ij} s - \sum_{(k,l)\neq(i,j)}f_{kl}
	||(\delta \mathbf{r}_k-\delta \mathbf{r}_l)_\text{tang}||^2\\
	&= f_{ij} s - C_{ij} s^2 + o(s^2)\text,
    \end{aligned}
    \label{eqn:cij}
\end{equation}
where $f_{kl}>0$ is the magnitude of the force between
$k$ and $l$ in the original packing, and $(\mathbf{v})_\text{tang}$ denotes the component
of the vector $\mathbf{v}$ in the plane tangent to the spheres $k$ and $l$ in the original packing \cite{wyart}.
At some value of $s=s_c$, a pair that was previously separated
by a gap will come into contact, and the motion is unphysical beyond this terminal point.
At a value $s^*\sim f_{ij}/C_{ij}$, the volume will start to decrease beyond its original value.
The packing is considered stable if for every single-contact-breaking motion, the volume at the
terminal point $s_c$, is increased compared to the original volume, i.e., if $s_c$ is of the order of $s^*$ or smaller
for all contacts. If violated at all,
this stability criterion will be violated by a contact whose force is of the same order
as the weakest
contact, and the gap closed will have a width of the same order as the smallest gap.
We will assume that the average number of non-contact-neighbor centers at a distance
of $r=1+\xi$ or less from the center of a randomly chosen sphere is
\begin{equation}
    Z_\text{nc}(\xi) \simeq A_d \xi^{1-\gamma}\text{ for }0<\xi\ll 1\text,
    \label{eq:qua}
\end{equation}
where $A_d$ is the dimension-dependent amplitude of the quasicontact singularity.
Similarly, we assume that the probability density of the force at a randomly chosen contact
is given by 
\begin{equation}
    P(f)\simeq f^\theta/\langle f\rangle^{1+\theta}\text{ for }0<f\ll\langle f\rangle\text,
    \label{eq:weak}
\end{equation}
where $\langle f\rangle$ is the average contact force and is proportional to the
externally applied pressure $p$. Then the smallest gap is of size roughly such that
$\tfrac{1}{2}N [Z(\xi_\text{min})-2d] = 1$, and so $\xi_\text{min} \sim (A_d N)^{-1/(1-\gamma)}$.
Similarly, as the total number of contacts is $dN$, the weakest contact is
roughly of strength $f_\text{min} \sim (dN)^{-1/(1+\theta)}\langle f\rangle$.
We assume, as in Ref.\ \cite{wyart}, that all single-contact-breaking motions extend to the entire system, and that each sphere moves by a
distance proportional to $s$ along each coordinate in a roughly uncorrelated way.
We therefore have that $C_{ij}\sim  d^2 N\langle f\rangle$, irrespective
of the strength of the contact broken. Therefore, for stability, we require that $s_c\sim\xi_\text{min}\sim (A_d N)^{-1/(1-\gamma)}$
be of the same order or smaller order than $s^*\sim d^{-1/(1+\theta)-2} N^{-1/(1+\theta)-1}$. From the dependence
on $N$, we obtain the inequality derived in Ref.\ \cite{wyart}, $\gamma \ge 1/(2+\theta)$.

\begin{figure*}[t]
    \centering
    \scalebox{1.}{
	\includegraphics[width=0.4\textwidth]{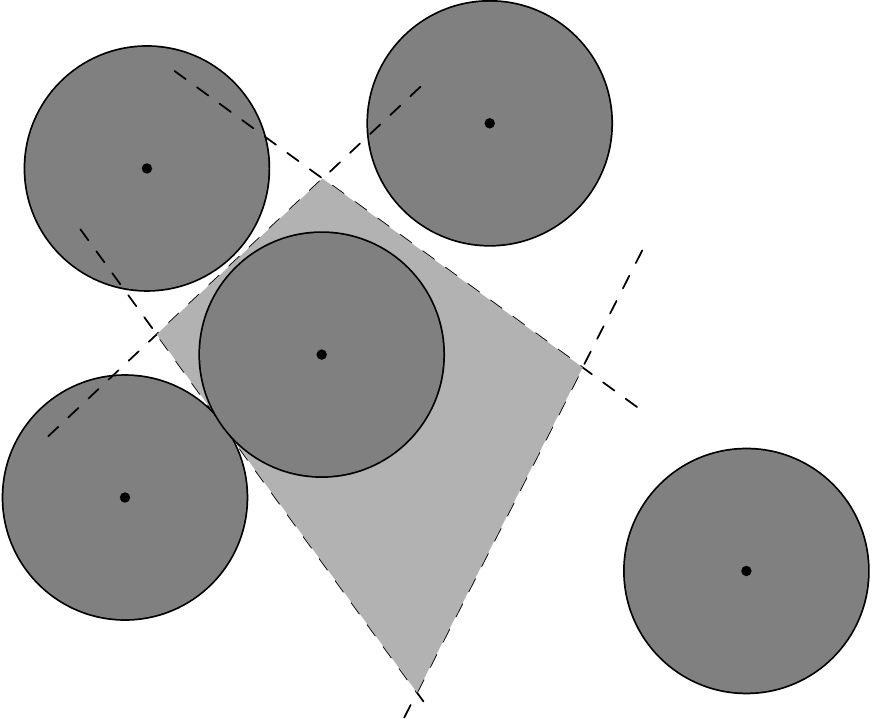}
	\hspace{5mm}
	\includegraphics[width=0.35\textwidth]{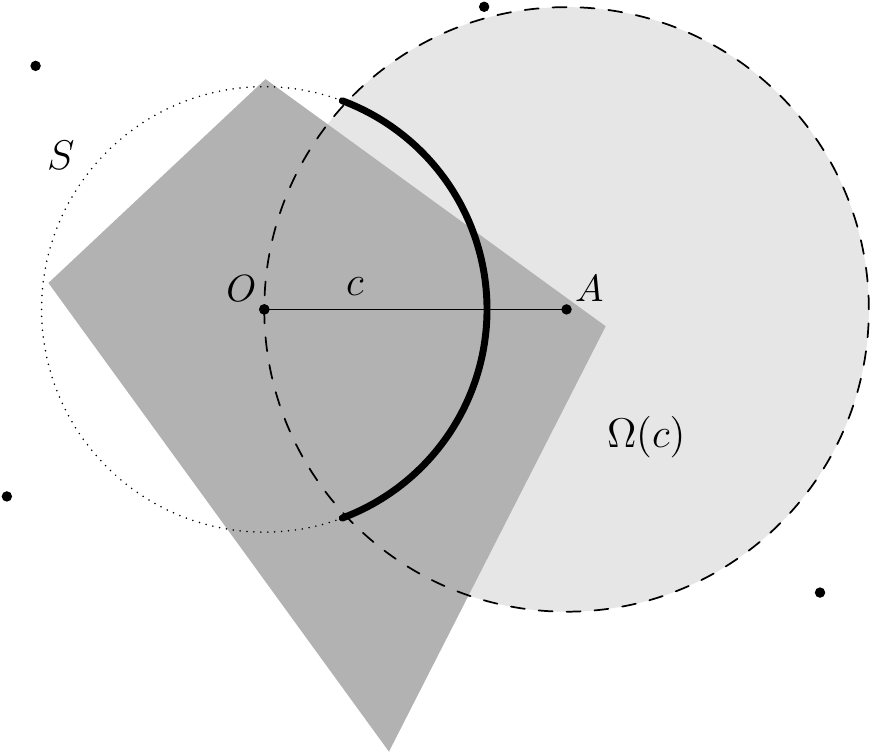}
    }
	\caption{Calculation of the (expected) volume of the Voronoi region around an arbitrary sphere.
	    The Voronoi region (medium gray) is the set of all points that are closer to the center of a given sphere
	    than to the center of any other (left). Point $A$ is in the Voronoi region of the sphere with center $O$
	    if and only if the (light gray) spherical region centered at $A$ and such that $O$ is on its
	    boundary has no sphere centers in its interior (right). Since the probability that this region
	    is empty depends only on the distance $c=|AO|$, we denote the region by $\Omega(c)$. The fraction
	    of a spherical surface $S$ (dotted line) of radius $r$ centered at $O$ that intersects $\Omega(c)$
	    (thick line) is denoted by $f(\tfrac{r}{2c})$.
	}
    \label{fig:vor}
\end{figure*}

Recall that mechanical stability required more contacts than degrees of freedom, and
the fact that MRJ packings lie at the margin between instability and stability implied
that this criterion would be satisfied with equality. Here, too, we may predict that
the inequalities we derive as criteria for dynamical stability will be satisfied with
equality, as a sign of marginal stability.
From the dependence on $d$, we have that the amplitude of the quasicontact singularity $A_d$ must increase
as $d^{\tilde{\nu}}$ with ${\tilde{\nu}} \ge (1 - \gamma)(3 + 2 \theta)/(1 + \theta)$, with equality under the
assumption of marginal stability. If we assume equality in both cases, we obtain ${\tilde{\nu}} = 2-\gamma$.
Note that we assume that the exponents $\gamma$ and $\theta$ are independent of dimension. We do not
know of any arguments from first principles that justify this assumption, but it is supported by
simulation results \cite{charbcorwin,iel}, and theoretical calculations in the limit $d\to\infty$ \cite{charbzamp}.

In a later paper, Lerner, D\"uring, and Wyart also analyzed the case of contacts that are weak because
their contact-breaking motion does not extend to the entire system. We address that case in Appendix A,
where we find that $\tilde{\nu}=2-\gamma$ also in the case that this type of contacts dominates.
As typical values for the power-law exponents seem to be not much different from
$\gamma\approx0.4$ and $\theta\approx0.3$ \cite{donev05,silbert06,charbcorwin},
we expect the number of quasicontacts to rise much faster
as a function of dimension than the number of contacts does. The mechanical description
of a packing depends only on the force-carrying contacts, not quasicontacts.
However, from a geometric point of view, contacts and quasicontacts are nearly indistinguishable. Therefore,
the number of quasicontacts, not only the number of mechanical contacts, is likely to be a key determinant of the
structure of MRJ packings in high dimensions.

\section{Effect of quasicontacts on density}\label{sec:dens}

To study the effect of quasicontacts, we will revisit a calculation of Jin et al. \cite{jin},
in which quasicontacts are originally neglected. Specifically, we will introduce their
presence into the calculation. The original calculation, presented in Ref.\ \cite{jin}, was
inspired by an earlier calculation \cite{song}
limited to three dimensions, and derives the density of a disordered packing in $d$ dimensions
based on the expectation value of the Voronoi region surrounding an arbitrary sphere.
The calculation is based on two assumptions: (i) a simplified form for the pair correlation function,
\begin{equation}
    g(r) = \Theta(r-1) +\frac{z \delta(r-1)}{\rho S_{d-1}}\text,
    \label{eq:grjin}
\end{equation}
where $z=2d$ is the average contact number, $\Theta$ is the Heaviside step function,
$\delta$ is the Dirac delta function, $S_{d-1}$ is the $(d-1)$-dimensional surface area of the
unit sphere, and $\rho=N/V$ is the number density of the packing; and (ii) after choosing
the origin to be the center of the sphere whose Voronoi region is to be calculated,
the centers of all other spheres are assumed to be uncorrelated with each other,
distributed according to a random
process whose radius-dependent density is given by $\rho g(r)$
\footnote{Despite the title of Ref.\ \cite{jin}, the calculation does not use the Edwards ensemble as an assumption.}.
The two assumptions are motivated by the principle of decorrelation in high dimensions,
which posits that as the spatial dimension increases, unconstrained
spatial correlations tend to vanish \cite{disorderwins}. Indeed, the features
of the pair correlation function away from $r=1$ appear to flatten out as the dimension
increases (see, e.g., Figure 7 of Ref.\ \cite{skoge}). However, the quasicontact
divergence, which is neglected in the form of $g(r)$ given in \ref{eq:grjin},
remains prominent.

The calculation proceeds by noting that a point at a distance $c$
from the origin is within the Voronoi region if and only if the
spherical region of radius $c$ surrounding that
point does not contain any sphere centers in its interior (see \ref{fig:vor}, where
this region is denoted as $\Omega(c)$).
The assumption (ii) gives rise to the rule that the probability that a region of space $\Omega$ contains
no sphere centers is given by
\begin{equation}
    P(N_\Omega=0) = \exp\left(-\langle N_\Omega\rangle\right)\text,
    \label{eq:pexp}
\end{equation}
where $N_\Omega$ is the number of sphere centers in $\Omega$.
This \textit{ansatz} ignores all correlations between different spheres.
In reality, the hard-core repulsion interaction implies a negative correlation between different spheres
\cite{torqvoid}, and so \ref{eq:pexp} should be taken as an upper bound in the two-sided inequality
\begin{equation}
    1 - \langle N_\Omega\rangle< P(N_\Omega=0) < \exp\left(-\langle N_\Omega\rangle\right)\text,
    \label{eq:pbnd}
\end{equation}
where the left-hand side represents perfect anti\-cor\-rela\-tion and the right-hand side represents
no correlations. Note that the right-hand side is a particularly poor approximation
of $P(N_{\Omega(c)}=0)$ when $\tfrac12 < c \le \sqrt{1/3}$. In that case, the left-hand
side is actually exact, since $\Omega(c)$ can contain at most one sphere center.

In the limit of very high dimensions, Jin et al.\ are able to show that,
given their assumptions, the density of the packing is given asymptotically
by $\varphi \sim d^\nu 2^{-d}$ with $\nu=1$.
Indeed, it has been shown that this density is a rigorous upper bound for any
packing that possesses a $g(r)$ specified by \ref{eq:grjin} \cite{disorderwins}.
In Appendix B, we replace the
assumption (i) with (i') a somewhat less simplified form for the pair correlation
that includes both contacts and quasicontacts:
\begin{equation}
    g(r) = \Theta(r-1) \left(1+\frac{(1-\gamma)A_d (r-1)^{-\gamma}}{\rho S_{d-1} r^{d-1}}\right)
    +\frac{z \delta(r-1)}{\rho S_{d-1}}\text,
    \label{eq:grq}
\end{equation}
where $A_d\sim d^{\tilde{\nu}}$ is the amplitude of the quasicontact singularity.
We show in Appendix B that under this modified assumption,
$\nu = \max(1,\, \tilde{\nu}+\gamma-1)$, where the first argument
of the maximum represents the effect of contacts, the second represents the effect
of quasicontacts, and the larger of the two dominates. Since we have seen that
marginal stability implies that $\tilde{\nu}=2-\gamma$, we have the unexpected
result that in marginally stable packings, neither contacts nor quasicontacts
dominate the effect of the other on the scaling of the density derived
in this calculation.

The presence of
quasicontacts does affect the prefactor of the asymptotic form of the density.
The calculation of Jin et al. predicts $2^d \varphi \approx 1.33 d$. Using
a value of $\gamma=0.42$, the calculation of Appendix B gives
$2^d \varphi \approx (1.33 + 1.00 c) d$, where $A_d \approx c d^{2-\gamma}$. To approximate
the magnitude of the added term, we estimate $c\approx 3.7$ by fitting $A_d=c d^{1.58}$
to values of $A_d$ measured in numerical simulations in dimensions $3-10$ \cite{charbcorwin}.
Therefore, it appears that the quasicontact contribution is considerably more
significant than the contact contribution in this calculation. It also helps
close the large gap that has been present between the prediction of Jin et al.
and other theoretical predictions that give the same scaling exponent but a much larger
prefactor, such as the prediction $2^d \varphi_\text{th} \approx 6.26 d$ of Ref.\ \cite{parisi10}.
Note that in cases in which the dynamic stability
criteria of \ref{sec:mrj} are satisfied with inequality rather than equality,
quasicontacts are predicted to dominate.
Note also that if the assumption (ii) is reinterpreted as an upper bound
as in \ref{eq:pbnd}, then the calculated density can be seen as a lower bound,
$\nu\ge\max(1,\, \tilde{\nu}+\gamma-1)$.

\section{Quasicontacts in lattice packings}\label{sec:bravais}

While the presence of quasicontacts turns out not necessarily to affect the
asymptotic scaling exponent of the density in MRJ packings, 
we show next that it does likely have an effect in marginally stable jammed
Bravais lattice packings.
Bravais lattice packings are periodic packings of spheres with one sphere
per unit cell. Recently, we explored an ensemble of jammed (mechanically stable)
Bravais lattices generated by a sequence of compressions in dimensions $d=15-24$ \cite{statlatt,marcotte,iel}.
Almost all of these lattices were found to possess the bare minimum number
of contacts required for mechanical stability of a Bravais lattice, namely $z=d(d+1)$,
and the pair correlation and force distribution of these jammed lattices exhibited power-law tails.
These findings suggest that these lattices are marginally stable with respect to both
static and dynamics stability, and should therefore be regarded as
an analog of MRJ packings in the context of Bravais lattices.
Bravais lattices with the minimal contact number $z=d(d+1)$ are hyperstatic, since $z>2d$, and
indeed they can remain mechanically stable even after an extensive number
of contact constraints are relaxed. However, if any contact constraint is relaxed
together with all of its periodic images, then the system loses its mechanical
stability. Therefore, with the idea of considering a single unit cell as our
system and the removal of one constraint entailing the removal of its images in
all other unit cells, we will refer to these lattices as isostatic in this specific
context, despite the fact that the packing as a whole is hyperstatic \cite{iel}. 

The power-law tails of the pair correlation and force distribution function 
were found to have dimensionally-independent exponents $\gamma=0.314\pm0.004$ and $\theta=0.371\pm0.010$ 
respectively, and The prefactor of the quasicontact
singularity was found to grow as a power law $A_d\sim d^{\tilde{\nu}}$,
with $\tilde{\nu}=3.30\pm0.05$ \cite{iel}. Before considering the effect of quasicontacts on the
typical density of these lattices, we begin by translating the calculation of
Ref.\ \cite{wyart} to the context of Bravais lattices in order to derive a stability criterion
in terms of the exponents $\gamma$, $\theta$, and $\tilde\nu$.

The problem of lattice sphere packing can be formulated as the problem of determining a symmetric, positive
definite matrix $G$ of minimal determinant, such that $\langle\mathbf{n},G\mathbf{n}\rangle\ge1$
for all nonzero integer vectors $\mathbf{n}\in\mathbb{Z}^d$ \cite{SPLAG}. The centers of the spheres then
take the positions $M\mathbf{n}$, where $M$ is some matrix such that $G=M^T M$, and
the density is $\varphi=\tfrac{1}{d} 2^{-d} S_{d-1} (\det G)^{-1/2}$. The jammed packings occur as locally optimal
solutions of this optimization problem, and such lattices are known as \textit{extreme} lattices \cite{perfect}.
The minimum number of contacts in an extreme lattice
is $d(d+1)$, corresponding to equality between the number of degrees of freedom in $G$,
$\tfrac{1}{2}d(d+1)$, and the number of active constraints
(the constraints $\langle\mathbf{n},G\mathbf{n}\rangle\ge1$ and $\langle-\mathbf{n},G(-\mathbf{n})\rangle\ge1$
are equivalent) \cite{perfect}. Extreme lattices with this minimum
contact number are called isostatic lattices \cite{iel} or \textit{minimally extreme} lattices \cite{kalluspack}.
We label a set of integer vectors responsible for all inequivalent
contacts as $\mathbf{n}_k$, $k=1,\ldots,d(d+1)/2$. From the method of Lagrange multipliers we
have that at a local minimum 
\begin{equation}
    (\det G) G^{-1} = \sum f_k \mathbf{n}_k \mathbf{n}_k^T\text,
    \label{eqn:force}
\end{equation}
where the contact forces $f_k$ are positive. The average force
is $\langle f\rangle = 2(\det G)/(d+1)$ \cite{iel}.

For any contact $\mathbf{n}_i$ in a minimally extreme lattice, there is a motion
$\delta G_i(s) = s G'_i$, such that
$\langle \mathbf{n}_k, \delta G_i(s) \mathbf{n}_k\rangle = s \delta_{ik}$ for all $k=1,\ldots,d(d+1)/2$.
This motion can continue until at some $s_c$, there is an integer vector $\mathbf{n}_c$ that does
not correspond to an original contact and becomes a contact $\langle \mathbf{n}_c, (G+sG'_i)\mathbf{n}_c\rangle=1$.
The identity $\det(1+X) = \exp(\sum_{n=1}^\infty\tfrac{(-1)^{n+1}}{n} \trc X^n)$ allows us to expand the determinant of $G+sG'_i$ 
as a power series:
\begin{equation}
    \begin{aligned}
	\frac{\det (G+sG'_i)}{\det G} 
	=&1 + s\trc G^{-1} G'_i \\
	+\tfrac{s^2}{2} \left[(\trc G^{-1} G'_i)^2 -\right.
	&\left.\trc G^{-1} G'_i G^{-1} G'_i \right] + o(s^2)\text.
    \end{aligned}
    \label{eqn:dV}
\end{equation}
We use \ref{eqn:force} to obtain
\begin{equation*}
    \begin{aligned}
	&(\det G)(\trc G^{-1} G'_i) =
	\sum_k f_k \langle \mathbf{n}_k, G'_i \mathbf{n}_k\rangle = f_i\text, \\
	&(\det G)^2(\trc G^{-1} G'_i G^{-1} G'_i) =
	\sum_{j,k} f_j f_k \langle \mathbf{n}_j, G'_i \mathbf{n}_k\rangle^2\text.
    \end{aligned}
    \label{eqn:tr1} 
\end{equation*}
Therefore, the change in the determinant of $G$ along the
single-contact-breaking motion is given by $\delta(\det G) = f_i s - C_i s^2 +o(s^2)$, where
\begin{equation}
    C_i = \frac{1}{2\det G} \sum_{j\neq k} f_j f_k \langle \mathbf{n}_j, G'_i \mathbf{n}_k\rangle^2\text.
\end{equation}
The typical magnitude of $C_i$ is given by $C_i\sim d^4 (\det G)^{-1} \langle f\rangle^2$.
For stability, we need $s_c$, the maximum physical extent of the motion associated with breaking a contact
carrying a force of the order of the weakest force, to be smaller than the extent $s^*$ for which $\delta(\det G)=0$.
Since we have $\sim d^2$ contacts, the smallest force is typically $f_\text{min}\sim d^{-2/(1+\theta)} \langle f\rangle$,
where $\langle f\rangle \sim (\det G)/d$. If the average number of non-contact lattice points at a distance
less than $r=1+\xi$ is $Z_\text{nc}(\xi) = A_d \xi^{1-\gamma}$ for $0<\xi\ll 1$, then the width of the smallest gap is of order
$\xi_\text{min} \sim A_d^{-1/(1-\gamma)}\sim d^{-{\tilde{\nu}}/(1-\gamma)}$. Therefore, for stability we require that
$s^*\sim (d^{-2/(1+\theta)} \langle f\rangle)/(d^{4} \langle f\rangle^{2} (\det G)^{-1}) \sim d^{(5+3\theta)/(1+\theta)}$
be of the same order or larger order than $d^{-{\tilde{\nu}}/(1-\gamma)}$. We have stability if
${\tilde{\nu}} \ge (1-\gamma)(5+3\theta)/(1+\theta)$, and marginal stability in the case of equality.
This inequality appears to be satisfied and nearly saturated based on the values of the
exponents derived from numerical simulations in dimensions $d=15-24$: 
$\tilde{\nu}=3.30\pm0.05$, $\gamma=0.314\pm0.004$, $\theta=0.371\pm0.010$
and $(1-\gamma)(5+3\theta)/(1+\theta)=3.06\pm0.02$.

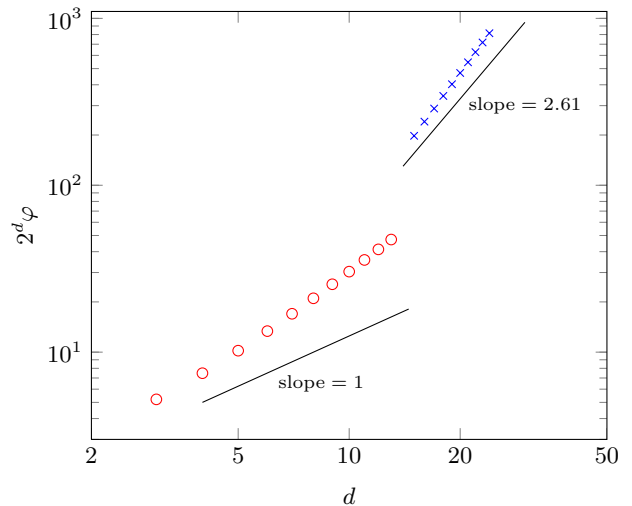
\begin{figure}[t]
    \centering
    \scalebox{1.}{
	\tikzsetnextfilename{dens}
	\begin{tikzpicture}
	    \begin{loglogaxis}[
		    xlabel={$d$}, ylabel={$2^d\varphi$},
		    xmin=2., xmax=50.,
		    ymin=3, ymax=1100,
		    y label style={at={(0.05,0.5)}},
		    xticklabels = {$2$,$5$,$10$,$20$,$50$},
		    xtick={2,5,10,20,50}
		]
		\addplot[mark=o, color=red, only marks] table[x index=0,y index=1] {charbonneau-phi.dat};
		\addplot[mark=x, color=blue, only marks] table[x index=0,y index=1] {phid.dat};
		\node[] at (axis cs:8.4,6.5) {\scriptsize $\text{slope}=1$};
		\addplot[] coordinates { (4,5) (14.5,18.125)};
		\node[] at (axis cs:30,300) {\scriptsize $\text{slope}=2.61$};
		\addplot[] coordinates { (14,130) (29.96,946.94)};
	    \end{loglogaxis}
	\end{tikzpicture}
    }
	\caption{Red circles: densities of disordered jammed packings as a function of dimension
	    from Ref.\ \cite{charbcorwin}. Blue crosses: densities of jammed Bravais lattice
	    packings from Ref.\ \cite{iel}. The black lines illustrate the scaling exponent
	    predicted as a lower bound by the calculation of Appendix B.
	}
    \label{fig:dens}
\end{figure}

\begin{figure}[t]
    \centering
    \scalebox{1.}{
	\tikzsetnextfilename{iso8f}
	\begin{tikzpicture}
	    \begin{axis}[
		    xlabel={$f/\langle f\rangle$}, ylabel={$\langle f\rangle P(f)$},
		    xmin=0, xmax=3.,
		    ymin=0, ymax=1.1,
		    y label style={at={(0.05,0.5)}},
		    domain=0:3.
		]
		\addplot[mark=triangle*, mark size=1., color=green, only marks] table[x index=0,y index=1] {iso8f.dat};
		\addplot[mark=*, mark size=0.5, color=purple, only marks] table[x index=0,y index=2] {iso8f.dat};
		\addplot[] {exp(-1.74758 + 3.0149*x - 1.58905*x^2)};
	    \end{axis}
	\end{tikzpicture}
    }
	\caption{
	    The distribution of contact forces over all contacts of all minimally extreme eight-dimensional lattices (green triangles)
	    and in the 200 densest lattices (purple circles). The line shows the best fit truncated Gaussian as a guide for the eye.
	}
    \label{fig:iso8f}
\end{figure}
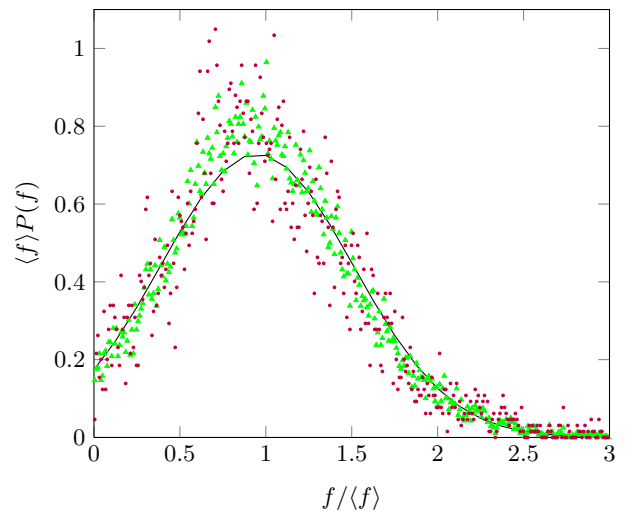

Since the principle of decorrelation applies also to lattice configurations
\cite{decorr, iel}, the assumptions of the calculation of Jin et al.\ should
apply for lattices in high dimensions to the same extent that they do for MRJ packings.
For lattices with the minimum number of contacts required for mechanical stability,
the contact number is $z=d(d+1)$. Therefore the
calculation yields the density estimate (or lower bound, if the second
assumption is interpreted as a bound) $\varphi\sim d^\nu 2^{-d}$,
where $\nu = \max(2,\tilde{\nu}+\gamma-1)\approx2.61$.
In this case the second term
of the maximum dominates and the quasicontact abundance is predicted
to have a significant effect on the asymptotic scaling of the density.
However, as we observe in Appendix B, the effect of quasicontacts is
only expected to dominate the effect of contacts in relatively high
dimensions. The calculation
suggests that the former overtakes the latter around $d\approx5000$,
much higher than the dimensions explored thus far in simulations.

With this observation in mind, we should be careful in using
low-dimensional data to extract high-dimensional scaling.
However, it seems clear from the data that the densities
of both MRJ packings and jammed Bravais lattice packings increase at a
much higher rate than predicted by our calculation
(see \ref{fig:dens}). It is reasonable
to conclude that the assumption (ii) neglects important anti-correlations
in these packings. Nevertheless, the interpretation of our calculation
as a lower bound is sound, since it is not expected that a less simplified
approximation of $g(r)$ than that of assumption (i') would yield a
dramatic decrease in the predicted density.

\section{Breakdown of the Edwards ensemble \textit{ansatz}}

Jammed lattice packings also provide a useful system for studying Edwards
statistical mechanics. Edwards proposed to investigate granular materials by considering ensembles
that bear a formal resemblance to classical thermodynamic ensembles but replace
energy conservation with volume conservation \cite{edwards}. The Edwards canonical ensemble
gives a weight proportional to $\exp(-XV)$ to any mechanically stable configuration
occupying a volume $V$ and zero weight to non-stable configurations.
The compactivity $X$ takes the role of inverse-temperature in the classical
canonical ensemble.
For the case of lattices, the mechanically stable configurations correspond
to extreme lattices. Voronoi proposed an algorithm to completely
enumerate perfect lattices, a superset of extreme lattices, in a given dimension,
and this full enumeration has been performed in dimensions up to $d=8$ 
\cite{perfect8}. In eight
dimensions, there are 10,916 perfect lattices, out of which 858 are minimally
extreme \cite{extreme8}. We can therefore study the Edwards ensemble exactly for $d=8$.
We focus on the distribution of contact forces, since the distribution
of gaps is still far from continuous for $d=8$. In \ref{fig:iso8f}, we show
the distribution of contact forces in the Edwards ensemble for zero compactivity.
Note that there appears to be a finite probability density at $f=0$, in contrast with
the expected power-law behavior. We suspect that this is not only
the case for zero compactivity, but unfortunately the available sample
in $d=8$ is not large
enough to discern a clear-cut violation of the power-law behavior
for compactivities large enough so that only a small part of the
sample has statistical weight. However, we note that the violation appears to persist even
when looking at only the 200 densest minimally extreme lattices, but the statistical
significance is much lower due to larger fluctuations.

\section{Conclusion}

In this paper we have investigated from a few different directions the power-law singularities
of the pair correlation function near the contact distance and of contact force probability density
near zero force. These singularities arise in MRJ packings and in their recently identified
relatives, jammed Bravais lattice packings. In both cases, they appear to be signatures of
the fact that these configurations lie at the margin between overconstraint and underconstraint
in terms of dynamic stability, much like isostaticity is related to marginal mechanical stability.
As we have seen in the case of Bravais lattice packings, power-law tails in the distributions
of active constraints on the verge of becoming inactive and \textit{vice versa} might also arise
in other scenarios in which cost is minimized in a quenched frustrated system subject to many competing
inequality constraints.

We showed that quasicontacts, the pairs of spheres contributing to the power-law singularity in the
pair correlation function, become much more abundant at high dimensions than actual contacts. Since
from a geometric, rather than a mechanical, perspective, quasicontacts can be nearly indistinguishable
from contacts, they could have a significant effect on the geometric structure of MRJ packings
in high dimension. Indeed, we show that a calculation of the density based on a simplified model
of correlations predicts that contacts and quasicontacts both contribute to the leading term of
the asymptotic dependence of the density on dimension, which is calculated to be $\varphi \sim d^\nu 2^{-d}$,
with $\nu = 1$. In fact, it seems that the contribution due to quasicontacts is significantly
larger than that due to contacts.
Moreover, in the case of Bravais lattices, the calculation predicts that
the quasicontact contribution dominates asymptotically.

Finally, we showed that in the case of Bravais lattices of dimensions not too high, it is possible to
directly study the Edwards ensemble. We considered the distribution of contact forces in the Edwards ensemble
of Bravais lattices in $d=8$, but we were unable to reproduce a power-law behavior for near-zero forces.
As we have argued, the power-law behavior of weak contacts and quasicontacts is a consequence of dynamic effects,
an end product of an unstable system driven out of equilibrium to the edge of stability, and we should not
be surprised that a theory based on equilibrium methods positing only static stability does not reproduce it.

\textbf{Acknowledgements.} We thank \'Etienne Marcotte for fruitful
discussion and feedback.
Y.\ K.\ would like to thank Patrick Charbonneau,
Matthieu Wyart, and Francesco Zamponi for useful comments. 
S.\ T.\ was supported in part
by the Materials Research Science and Engineering Center
Program of the National Science Foundation under Grant No.
DMR-0820341 and by the Division of Mathematical Sciences
at the National Science Foundation under Award No. DMS-1211087.
This work was partially supported by a grant from the
Simons Foundation (Grant No. 231015 to Salvatore Torquato).

\appendix

\section{}

In the main text we considered the stability of a disordered packing with
respect to motions that break a weak contact and extend to the entire
packing. In Ref.\ \cite{lerner}, the authors demonstrated the existence
of contacts that lead to localized contact-breaking motions and claim
that such contacts dominate the distribution of weakest contacts.
As we showed in the main text, the contact force $f_{ij}$ is identified
with the first-order derivative $\lim_{s\to 0} p\delta V(s)/s$ of the
rate of work done upon opening the contact into a gap of width $s$,
maintaining all other contacts. Lerner et al.\ decomposed the contact force
$f_{ij}=b_{ij}w_{ij}$ into two factors: $b_{ij}$ measures how strongly
the contact-breaking motion couples to the bulk, i.e., the extent to
which a typical sphere in the packing moves in proportion to $s$; $w_{ij}$
measures how much the bulk motion tends to increase the volume of the packing,
i.e., how strongly it is coupled to the movement of spheres at the boundary in
the direction orthogonal to the boundary. The case treated in the main
text corresponds to the case in which the weak contact distribution is dominated
by contacts with $b_{ij}\sim1$ and $w_{ij}$ is distributed according to 
$P(w)\sim w^\theta$. In this appendix we treat the case in which $w_{ij}\sim1$
and $b_{ij}$ is distributed according to $P(b)\sim b^\theta$.

While spheres in the bulk move by distances $\sim b s$, where $b\ll N^{-1/2}$,
the number of spheres that move by distances $\sim s$ does not increase with
the system size.
However, because the average number of contacts around a sphere is $2d$,
the number of spheres involved in the localized motion will increase with
dimension as $\sim d$. Since the gap whose closing defines the physical
extent $s_c$ of the motion is likely to be far away from the contact being
opened, we have
$s_c\sim\xi_{\text{min}}/b_{ij}\sim\xi_{\text{min}}/(f_{ij}/\langle f\rangle)$,
where $\xi_\text{min}\sim (A_d N)^{-1/(1-\gamma)}$ is the smallest gap in the packing.
Meanwhile, $C_{ij} = \lim_{s\to 0} \sum_{(k,l)\neq(i,j)} f_{kl}
	||(\delta \mathbf{r}_k-\delta \mathbf{r}_l)_\text{tang}||^2/s^2$,
the coefficient that controls the second-order change in volume,
is dominated by the $\sim d$ spheres involved in the localized motion,
which move a distance $\sim s$. Therefore, the typical value
of $C_{ij}$ for these contacts is $\sim d \langle f\rangle$
and $s^*\sim f_{ij}/C_{ij} \sim d^{-1}(f_{ij}/\langle f\rangle)$. For contact forces
of the same order as the weakest contact,
$f_\text{min}\sim(dN)^{-1/(1+\theta)}$, stability requires
that $s_c\sim d^{-\tilde{\nu}/(1-\gamma)+1/(1+\theta)} N^{-1/(1-\gamma)+1/(1+\theta)}$
be of the same order or smaller order than $s^*\sim d^{-1-1/(1+\theta)} N^{1/(1+\theta)}$.
The conclusion is that the criteria for stability are
$\gamma \ge (1-\theta)/2$ and $\tilde{\nu}\ge(1-\gamma)(3+\theta)/(1+\theta)$.
If we assume equality in both cases, we obtain $\tilde{\nu}=2-\gamma$.
Therefore, no matter which kind of contacts dominate the
distribution of weakest contacts, we obtain the same prediction
for the exponent controlling the dimensional growth of the
abundance of quasicontacts.

\section{}

In this appendix, we derive an asymptotic expression for the density
of an MRJ packing in the limit of high dimensions by calculating
the expected volume of the Voronoi region associated with a typical
sphere in the packing. The calculation is based on two assumptions:
(i') a simplified form \ref{eq:grq} for the pair correlation that includes terms
corresponding to contacts, quasicontacts, and bulk spheres, and (ii)
a lack of significant correlations between spheres other than the pair correlations
involving the central sphere whose Voronoi volume is being calculated. This
latter assumption is expressed by the \textit{ansatz} \ref{eq:pexp}, which can alternatively
be interpreted as a bound \ref{eq:pbnd}.

The expected volume of the Voronoi region is given by
\begin{equation}
    \rho^{-1} = \langle v\rangle = \frac{2^{-d}S_{d-1}}{d} + \int_{c=\tfrac{1}{2}}^{\infty}  S_{d-1}c^{d-1} P(c) dc\text,
    \label{eq:vint}
\end{equation}
where $P(c)$ is the probability that a point at a distance $c$ from the origin
is inside the Voronoi region. Note that \ref{eq:vint} simply reflects the
additivity of the expectation value of the volume of intersection of the Voronoi region
with any region of space, regardless of correlations. Therefore, it is exact in every
dimension, provided a correct form of the function $P(c)$.

We use assumption (ii) to give an approximation of $P(c)$, the probability that
the region $\Omega(c)$ contains no sphere centers (see \ref{fig:vor}):
\begin{equation}\begin{aligned}
    P(c) &= \exp (-\langle N_{\Omega(c)}\rangle)\\
    &= \exp\left(-\int_{r=1}^{2c} r^{d-1} \rho g(r) S_{d-1} f(\tfrac{r}{2c}) dr\right)\text,
    \end{aligned}\label{eq:pcint}
\end{equation}
where $\Omega(c)$ is a spherical region of radius $c$ whose center is a distance $c$ from
the origin, and $0\le f(\tfrac{r}{2c})\le1$ is the fraction of a spherical surface of radius $r$
centered at the origin that intersects $\Omega(c)$. The fraction $f(x)$
of the sphere at a latitude of less than $\arccos x$ from the pole
is given by the integral
\begin{equation}
    f(x) = \frac{S_{d-2}}{S_{d-1}} \int_{t=x}^{1} (1-t^2)^{\tfrac{d-3}{2}}dt\text.
    \label{eq:ft}
\end{equation}

We may decompose $P(c)=P_B(c)P_C(c)P_Q(c)$ into independent terms
corresponding to spheres in the bulk, contacts, and quasicontacts respectively.
We have
\begin{equation}\begin{aligned}
    \log P_B(c) &= -\rho S_{d-1} \int_{r=1}^{2c} r^{d-1} f(\tfrac{r}{2c}) dr\\
    &=-\tfrac{\rho}{d} S_{d-2} \int_{t=1/2c}^1(1-t^2)^{\tfrac{d-3}{2}}(2^d t^d c^d -1)dt\\
    &=-\tfrac{\rho}{d} S_{d-1} [c^d(1-f(1-\tfrac{1}{2c^2}))  \\
    &\quad + \tfrac{c S_{d-2}}{(d-1)S_{d-1}} (1-\tfrac{1}{4c^2})^{\tfrac{d-1}{2}} - f(\tfrac{1}{2c})]\text;
    \end{aligned}\label{eq:pbc}
\end{equation}

\begin{equation}
    \log P_C(c) = -z f(\tfrac{1}{2c})\text;
\end{equation}
and
\begin{equation}\begin{aligned}
    \log P_Q(c) &= -(1-\gamma)A_d\int_{r=1}^{2c} (r-1)^{-\gamma} f(\tfrac{r}{2c}) dr\\& = - A_d h_\gamma(\tfrac{1}{2c})\text,
    \end{aligned}\label{eq:pqc}
\end{equation}
where
\begin{equation}
    h_\gamma(x) = \frac{S_{d-2}}{S_{d-1}} \int_{t=x}^{1}(1-t^2)^{\tfrac{d-3}{2}} (\tfrac{t}{x}-1)^{1-\gamma}dt\text.
\end{equation}
Note that $f(x)=h_1(x)$.

We let $w=(2^d\varphi)^{-1}=d/(\rho S_{d-1})$ and obtain the equation which determines $w$ implicitly,
\begin{equation}\begin{aligned}
	1 &= \tfrac{2^{-d}}{w} + \tfrac{d}{w} \int_{c=\tfrac{1}{2}}^{\infty} c^{d-1} e^{-\tfrac{c^d}{w}+H(c;w)}dc\\
	&=\tfrac{2^{-d}}{w} + \int_{y=\tfrac{2^{-d}}{w}}^{\infty} e^{-y+H(y^{1/d}w^{1/d};w)}dy\text,
    \label{eq:sfy}\end{aligned}
\end{equation}
where 
\begin{equation}\begin{aligned}
    H(c;w) &= \frac{c^d}{w} f(1-\tfrac{1}{2c^2}) - \frac{cS_{d-2}}{(d-1)wS_{d-1}}(1-\tfrac{1}{4c^2})^{\tfrac{d-1}{2}}\\
    &\quad+(\tfrac{1}{w} -z) f(\tfrac{1}{2c}) - A_d h_\gamma (\tfrac{1}{2c})\text.
    \end{aligned}
\end{equation}
For large $d$, we expect that $2^{-d}/w\to0$ and we may write $\int_{y=0}^{\infty}e^{-y}e^H(y^{1/d}w^{1/d};w)dy\approx1$. Using
the fact that $H(c;w)$ behaves nicely, we conclude that $w$ is such that $H(y^{1/d}w^{1/d};w)\approx0$ when $y\approx 1$.

We now turn to the task of approximating the functions $h_\gamma(x)$ and $f(x)=h_1(x)$ when $d\gg1$ and $x\nll1$. We write
$(1-t^2)^{\tfrac{d-3}{2}}$ as $\tfrac{d-3}{2}\exp(\log (1-t^2))$ and use Laplace's
method by expanding the argument of the exponential function about the point where it takes its largest 
value in the integration domain, which in this case is the lower bound $t=x$:
\begin{equation}\begin{aligned}
    \log (1-t^2) &\approx \log(1-x^2) +\frac{x^2}{1+x^2}\\&- \frac{1+x^2}{(1-x^2)^2}\left[(t-x)+\frac{x(1-x^2)}{1+x^2}\right]^2\text.
    \end{aligned}
\end{equation}
If we let $s = (d-3)^{\tfrac12}(1+x^2)^{\tfrac12}(t-x)/(1-x^2)$, we obtain an integral of the form
\begin{equation}\begin{aligned}
	h_\gamma(x)&\approx\frac{S_{d-2}(1-x^2)^{\tfrac{d-3}{2}+2-\gamma}} 
	{S_{d-1}(d-3)^{\tfrac{2-\gamma}{2}}(1+x^2)^{\tfrac{2-\gamma}{2}}x^{1-\gamma}}\\
		    &\quad\times e^{\tfrac{s_0^2}{2}}\int_{s=0}^{\infty} e^{-\tfrac{(s+s_0)^2}{2}}s^{1-\gamma}ds\text,
    \end{aligned}
\end{equation}
where $s_0 = (d-3)^{\tfrac12}x/(1+x^2)^{\tfrac12}$. The last integral can be asymptotically expanded in
powers of $1/s_0$ by performing a change of variables, expanding the resulting binomials,
and integrating term by term. We only require the first term of the expansion:
\begin{equation}\begin{aligned}
    &e^{\tfrac{s_0^2}{2}}\int_{s=0}^{\infty}e^{-\tfrac{(s+s_0)^2}{2}}s^{1-\gamma}ds \\
    &=\tfrac12 s_0^{2-\gamma}\int_{u=0}^{\infty}e^{-\tfrac{s_0^2}{2}u}
    \left((1+u)^{\tfrac12}-1\right)^{1-\gamma} (1+u)^{-\tfrac12}du\\
    &\approx \tfrac12 s_0^{2-\gamma}\int_{u=0}^{\infty}e^{-\tfrac{s_0^2}{2}u}(\tfrac12u)^{1-\gamma}du=s_0^{\gamma-2}\Gamma(2-\gamma)
    \end{aligned}
\end{equation}
We therefore obtain the asymptotic forms for $h_\gamma(x)$ and $f(x)=h_1(x)$:
\begin{align}
    h_\gamma(x) &\approx \frac{S_{d-2}\Gamma(2-\gamma)(1-x^2)^{\tfrac{d+1}{2}-\gamma}}
    {S_{d-1} (d-3)^{2-\gamma} x^{3-2\gamma}}\\
    f(x) &\approx \frac{S_{d-2}(1-x^2)^{\tfrac{d-1}{2}}}{S_{d-1} (d-3) x}\text.
\end{align}
The latter of these agrees with the asymptotic form derived in Ref.\ \cite{jin}.
Finally, then, we have that $H(w^{1/d};w)\approx0$, when
\begin{equation}\begin{aligned}
    &\frac{1}{2-w^{-\tfrac{2}{d}}}-\frac{d-3}{2d-2}+1-wz\\
    &\quad-wA_d\Gamma(2-\gamma)\frac{(1-\tfrac14w^{-\tfrac{2}{d}})(2w^{\tfrac1d})^{2-2\gamma}}{(d-3)^{1-\gamma}}\approx0\text.
    \end{aligned}
\end{equation}
So if $A_d\sim d^{\tilde{\nu}}$ and $z\sim d^{\nu_z}$, then $w\sim d^{-\nu}$, where $\nu=\max(\nu_z, \tilde{\nu}+\gamma-1)$.
If the first argument of the maximum dominates we have $\varphi=2^{-d}/w\approx2^{-d}(2z/3)$,
and if the second dominates we have $\varphi\approx2^{-d}(2^{1-2\gamma} \Gamma(2-\gamma)A_d d^{\gamma-1})$.

For jammed minimally extreme Bravais lattices, where $z=d(d+1)$,
numerical data for $d=15-24$ suggests that $A_d\approx (1.43\times10^{-3})d^{3.30}$
and $\gamma\approx0.314$. These values imply that in the limit of high dimensions, the density
estimate given by this calculation is dominated by the effect of quasicontacts. However,
note that $2^{1-2\gamma}\Gamma(2-\gamma)A_d d^{\gamma-1}\gg\tfrac23z$ only when
$d\gg5000$. For lower dimensions, the effect of contacts dominates or the two effects are comparable.

\bibliography{iel}

\end{document}